\documentclass{aa}
\usepackage[varg]{txfonts}
\usepackage{longtable}
\usepackage{lscape}
\usepackage{graphicx}
\newcommand{\kmprs}  {\mbox{\rm km\,s$^{-1}$}}
\newcommand{\feh} {\mbox{\rm [Fe/H]}}

\newcommand{\xh} {\mbox{\rm [X/H]}}

\newcommand{\xfe} {\mbox{\rm [X/Fe]}}
\newcommand{\cfe} {\mbox{\rm [C/Fe]}}
\newcommand{\ofe} {\mbox{\rm [O/Fe]}}
\newcommand{\nafe} {\mbox{\rm [Na/Fe]}}
\newcommand{\mgfe} {\mbox{\rm [Mg/Fe]}}

\newcommand{\alfe} {\mbox{\rm [Al/Fe]}}

\newcommand{\sife} {\mbox{\rm [Si/Fe]}}

\newcommand{\cafe} {\mbox{\rm [Ca/Fe]}}

\newcommand{\tife} {\mbox{\rm [Ti/Fe]}}
\newcommand{\crfe} {\mbox{\rm [Cr/Fe]}}
\newcommand{\mnfe} {\mbox{\rm [Mn/Fe]}}
\newcommand{\nife} {\mbox{\rm [Ni/Fe]}}
\newcommand{\cufe} {\mbox{\rm [Cu/Fe]}}
\newcommand{\znfe} {\mbox{\rm [Zn/Fe]}}
\newcommand{\yfe} {\mbox{\rm [Y/Fe]}}

\newcommand{\bafe} {\mbox{\rm [Ba/Fe]}}

\newcommand{\alphafe} {\mbox{\rm [$\alpha$/Fe]}}
\newcommand{\teff}  {\mbox{$T_{\rm eff}$}}
\newcommand{\Tc}  {\mbox{$T_{\rm cond}$}}

\newcommand{\logg}  {\mbox{{\rm log}\,$g$}}

\newcommand{\turb}  {\mbox{$\xi_{\rm turb}$}}

\newcommand{\CI} {\ion{C}{i}}
\newcommand{\OI} {\ion{O}{i}}

\newcommand{\MgI} {\ion{Mg}{i}}
\newcommand{\NaI} {\ion{Na}{i}}

\newcommand{\SiI} {\ion{Si}{i}}

\newcommand{\CaI} {\ion{Ca}{i}}

\newcommand{\TiI} {\ion{Ti}{i}}

\newcommand{\CrI} {\ion{Cr}{i}}

\newcommand{\MnI} {\ion{Mn}{i}}
\newcommand{\FeI} {\ion{Fe}{i}}

\newcommand{\NiI} {\ion{Ni}{i}}
\newcommand{\CuI} {\ion{Cu}{i}}
\newcommand{\ZnI} {\ion{Zn}{i}}

\newcommand{\YII} {\ion{Y}{ii}}

\newcommand{\BaII} {\ion{Ba}{ii}}

\def\ltsima{$\; \buildrel < \over \sim \;$}
\def\simlt{\lower.5ex\hbox{\ltsima}}
\def\gtsima{$\; \buildrel > \over \sim \;$}
\def\simgt{\lower.5ex\hbox{\gtsima}}
\begin{document}

\title{G\,112-43/44: A metal-poor binary star with a unique chemical composition and 
kinematics like the Helmi streams
\thanks{Based on observations made with the ESO Telescopes
at the La Silla Paranal Observatory under programmes 69.D0679, 70.D-0474, 76.B-0133,
and on observations made with the Nordic Optical Telescope
on La Palma.}}

\titlerunning{G\,112-43/44: A metal-poor binary star}

\author{P.E.~Nissen \inst{1} \and J.S.~Silva-Cabrera \inst{2} \and W.J.~Schuster \inst{3} }

\institute{Stellar Astrophysics Centre, 
Department of Physics and Astronomy, Aarhus University, Ny Munkegade 120, DK--8000
Aarhus C, Denmark  \email{pen@phys.au.dk}
\and CONACYT, Instituto de Astronomia, Universidad Nacional Aut\'{o}noma
de M\'{e}xico, AP 106, Ensenada 22800, BC, M\'{e}xico 
\and Instituto de Astronomia, Universidad Nacional Aut\'{o}noma
de M\'{e}xico, AP 106, Ensenada 22800, BC, M\'{e}xico}

\date{Received 18 March 2021 / Accepted 28 April 2021}

\abstract
{Correlations between high-precision elemental abundance ratios
and kinematics of halo stars provide interesting information about the formation and 
early evolution of the Galaxy.}
{Element abundances of G\,112-43/44, a metal-poor wide-orbit binary star with extreme kinematics, are revisited.}  
{High-precision studies of the chemical compositions of 94 metal-poor dwarf
stars in the solar neighbourhood are used to compare abundance ratios for G\,112-43/44 with ratios 
for stars having similar metallicity taking into account the effect of 
deviations from local thermodynamic equilibrium on the
derived abundances, and $Gaia$ EDR3 data are used to compare the kinematics.}
{The abundance ratios X/Fe of the two components of G\,112-43/44 agree within $\pm 0.05$\,dex for
nearly all elements, but there is a hint
of a correlation of the difference in \xh\ with elemental condensation temperature,
which  may be due to planet-star interactions. The Mg/Fe, Si/Fe, Ca/Fe, and Ti/Fe 
ratios of G\,112-43/44 agree with the corresponding ratios for  
accreted (Gaia-Enceladus) stars, but Mn/Fe, Ni/Fe, Cu/Fe, and Zn/Fe are significantly enhanced 
with $\Delta \, \znfe$ reaching 0.25\,dex. The kinematics show that 
G\,112-43/44 belongs to the Helmi streams in the solar neighbourhood and in view of this, 
we discuss if the abundance peculiarities of G\,112-43/44 can be explained 
by chemical enrichment from supernovae events in the progenitor dwarf galaxy of the Helmi streams. 
Interestingly, yields calculated for a helium shell detonation 
Type Ia supernova model can explain the enhancement of Mn/Fe, Ni/Fe, Cu/Fe, and Zn/Fe 
in G\,112-43/44 and three other $\alpha$-poor stars
in the Galactic halo with abundances from the literature, one of which have Helmi streams kinematics.
The helium shell detonation model predicts, however, also 
enhanced abundance ratios of Ca/Fe, Ti/Fe, and Cr/Fe in disagreement with the observed ratios.}
{}


\maketitle

\section{Introduction}
\label{introduction} 
High-precision studies of the chemical composition of metal-poor, high-velocity
stars in the solar neighbourhood
have provided interesting information about the formation and early evolution of the Milky Way.
Until 25 years ago, it was believed \citep[see review by][]{mcwilliam97}
that nearly all stars with $\feh < -1.0$ have a uniformly 
enhanced ratio between the abundance of $\alpha$-capture elements and iron  
relative to the corresponding ratio for the Sun, that is 
\alphafe\,\footnote{In this paper `$\alpha$' denotes the mean abundance of
Mg, Si, Ca, and Ti.} $\sim  +0.4$, as expected if core collapse supernovae (SNe) 
are the only source of chemical enrichment at low metallicities \citep{kobayashi06}.   
Work around the beginning of the 21st century revealed, however, that there are 
significant variations of \alphafe\ correlated with kinematical properties;
stars with high space velocities and retrograde moving stars were found to have
lower \alphafe\ values than prograde moving halo stars \citep{nissen97, hanson98, 
fulbright02, stephens02, gratton03, jonsell05, ishigaki10}. The differences in \alphafe\ are,
however, only in the order of 0.10 to 0.15\,dex, and with the precision obtained in
the cited papers, it was unclear if there is a continuum of \alphafe\ values at a given \feh\
or a bimodal distribution.

This question was answered by \citet{nissen10}, who in a
study of 94 high-velocity F and G dwarf stars reached a precision of 0.02\,dex for the differential values of 
\alphafe\ and showed that stars in the metallicity range $-1.4 < \feh < -0.7$ 
are distributed in two distinct populations, high-$\alpha$ and low-$\alpha$ halo stars.
These two populations are also well separated in other abundance ratios such
as \nafe , \nife , \cufe , and \znfe . Furthermore, the low-$\alpha$
stars tend to move on retrograde orbits and reach larger apo-galactic distances
than the high-$\alpha$ stars \citep{schuster12}. This and the decreasing
trend of \alphafe\ with increasing \feh , like in dwarf galaxies
\citep[see review by][]{tolstoy09}, suggest that the low-$\alpha$ population has been accreted
from dwarf galaxies in which Type Ia SNe start to contribute Fe at a relatively low metallicity 
because of a low star-formation rate. The high-$\alpha$ stars, on the other hand,
may have formed $in \, situ$ in a dissipative component with such a high star-formation rate 
that only Type II SNe contributed to the chemical evolution.

The findings of \citet{nissen10} have been confirmed for a larger sample of 
K giants with high-precision APOGEE abundances by \citet{hawkins15} and \citet{hayes18},
who added \alfe\ as a powerful  
discriminator between high- and low-$\alpha$ stars. 
Furthermore, \citet{gaia.collaboration18a}
showed that the colour-magnitude diagram of stars with high tangential velocities,
$V_{\rm t} > 200$\,\kmprs , has two separated
sequences in the turn-off region of which the blue sequence corresponds to low-$\alpha$ stars
and the redder sequence to high-$\alpha$ stars \citep{haywood18}. Another interesting result
from $Gaia$ DR2 data was obtained  by \citet{helmi18}, who showed that the low-$\alpha$ stars 
form an elongated structure in the Toomre velocity diagram
with a slightly retrograde mean rotation.
Based on this, they suggested that the low-$\alpha$ stars
are the debris of an accreted massive dwarf galaxy, named `Gaia-Enceladus'. 
The high-$\alpha$ population was explained as formed in a precursor to the thick disk and heated to halo
kinematics in connection with the Gaia-Enceladus merger.

In a comprehensive review, \citet{helmi20} argues that the large extension in metallicity 
and the small width in \alphafe\ of the low-$\alpha$ sequence (consistent with the measurement errors 
of the APOGEE abundances) make it unlikely that this population consists 
of stars from several small-mass dwarf galaxies. \citet{nissen11} find, however, significant
variations in \xfe\ among the low-$\alpha$ stars. At a given \feh , the variations in
\xfe\ for the high-$\alpha$ population  correspond to the measurement errors, but for the
low-$\alpha$ stars, the variations in \nafe , \mgfe , \nife , and \cufe\ are two to three times
larger. This can be explained if the low-$\alpha$ population has been accreted from
at least two dwarf galaxies with different star formation rates. Furthermore, according to
\citet{mackereth19}, APOGEE abundances provide evidence of systematic differences in
\mgfe , \alfe , and \nife\ for low-$\alpha$ stars having respectively high and low orbital eccentricity,
and other works \citep{matsuno19, koppelman19a, monty20}  suggest that high-energy
retrograde halo stars, belonging to the so-called Gaia-Sequoia event, stand out
by having lower \nafe , \mgfe , and \cafe\ ratios than the bulk of the Gaia-Enceladus stars.
Although these variations could be related to radial abundance gradients in the 
Gaia-Enceladus progenitor galaxy \citep{koppelman20}, they can also be explained if several 
dwarf galaxies contributed to the low-$\alpha$ population. Clearly, it would be interesting
to obtain additional high-precision abundances for metal-poor stars with different kinematics.

As a small contribution to this discussion, we present a detailed discussion of the metal-poor
binary star G\,112-43/44 that  belongs to the low-$\alpha$ population but has
enhanced abundance ratios of Mn/Fe, Ni/Fe, and Zn/Fe according to 
\citet{nissen10, nissen11}. In Sect. 2, we use $Gaia$ EDR3 data to show that G\,112-43/44 
belongs to one of the Helmi streams in the solar neighbourhood, and
in Sect. 3, we compare the abundances of the two components with other low-$\alpha$
stars having similar metallicity. In Sect. 4, we discuss how the derived [X/Fe] 
enhancements of G\,112-43/44 can be explained, and some conclusions are presented in Sect. 5.

\section{Kinematics}
\label{kinematics}
The components of G\,112-43/44, alias BD\,+00\,2058 A and B, are separated by 12 arcsec
on the sky and have visual magnitudes of $V_{\rm A} = 10.19$ and $V_{\rm B} = 11.17$
\citep{schuster93}. They were listed as 
a common proper motion pair in \citet{luyten79} and identified as a physical pair by
\citet{ryan92} on the basis of photometric distances and spectroscopic metallicities
from \citet{laird88}, $\feh = -1.44$ and $-1.46$, respectively. The physical connection is 
confirmed by $Gaia$ satellite observations as seen from Table \ref{table:data}, where data from
the Early Data Release 3 (EDR3) \citep{gaia.collaboration16, gaia.collaboration20}  
are given. As seen, the proper motions, radial velocities,
parallaxes, and the derived space velocities of the two components agree very well.
Further evidence for a physical
connection comes from the agreement of the chemical abundances of the two stars
as  discussed in Sect. \ref{discussion}.

\begin{table}
\caption[ ]{Kinematical parameters for G\,112-43 and G\,112-44 based on $Gaia$ EDR3 data.}
\label{table:data}
\centering
\setlength{\tabcolsep}{0.20cm}
\begin{tabular}{lcc}
\noalign{\smallskip}
\hline\hline
\noalign{\smallskip}
 Parameter   & G\,112-43 & G\,112-44 \\
\noalign{\smallskip}
\hline
\noalign{\smallskip}
$\alpha(2000.0)$            & $07\,43\,43.78$      & $07\,43\,43.90$  \\
$\delta(2000.0)$            & $-00\,04\,05.9$      & $-00\,03\,54.2$ \\
$\Delta \alpha$  [mas/yr]   & $-171.877 \pm 0.021$ & $-172.271  \pm 0.048 $  \\
$\Delta \delta$  [mas/yr]   & $-305.492 \pm 0.013$ & $-306.430  \pm 0.037 $  \\
RV [\kmprs]                 & $-82.69 \pm 0.33$    & $-82.64 \pm 0.66$   \\  
Parallax [mas]              & $5.595  \pm0.019 $   & $5.565  \pm 0.030 $  \\
Distance\,\tablefootmark{a} [pc]               & $178.7 \pm 0.6$      & $179.7 \pm 1.0$     \\
$U_{\rm LSR}$ [\kmprs]      & $136.7 \pm 0.3$      & $137.2 \pm 0.6$    \\
$V_{\rm LSR}$ [\kmprs]      & $-93.5 \pm 0.6$      & $-94.9 \pm 1.0$   \\
$W_{\rm LSR}$ [\kmprs]      & $-253.7 \pm 0.8$     & $-255.7 \pm 1.4$  \\
\noalign{\smallskip}
\hline
\end{tabular}
\tablefoot{
\tablefoottext{a} {Calculated as the inverse of the parallax, because the relative errors of the
parallaxes are less than 1\%, and by neglecting possible systematic errors in the parallaxes,
which are in the same order of size as the statistical errors \citep{lindegren20}.}
}
\end{table}

\begin{figure}
\centering
\includegraphics[width=8.5cm]{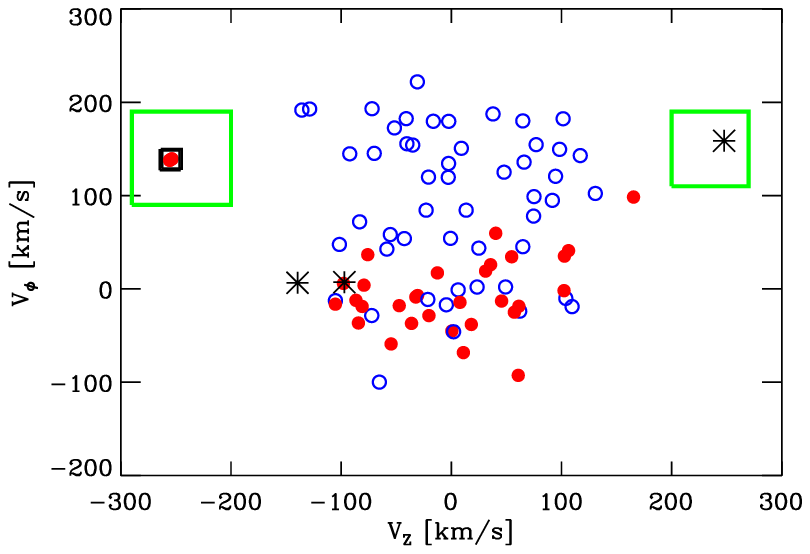}
\caption{$V_{\phi}$--$V_{Z}$  diagram for stars with $\feh > -1.4$ in \citet{nissen10}.
$V_{\phi}$ is the Galactic
rotational velocity component, $V_{\phi}$ = $V_{\rm LSR} + 232.8$\,\kmprs , and $V_{Z}$ is the
component perpendicular to the Galactic plane, $V_{Z}$ = $W_{\rm LSR}$. Stars from \citet{nissen10}
belonging to the
high-$\alpha$ sequence are shown with blue circles and those on the low-$\alpha$ sequence with
filled red circles. G\,112-43/44 is indicated with a black square. The three $\alpha$-poor halo stars
discussed in Sect. \ref{discussion} are shown with asterisks. The green
boxes indicate the location of the Helmi streams according to \citet{koppelman19b}.}
\label{fig:VphiVz}
\end{figure}

\begin{figure*}
\centering
\includegraphics[width=16cm]{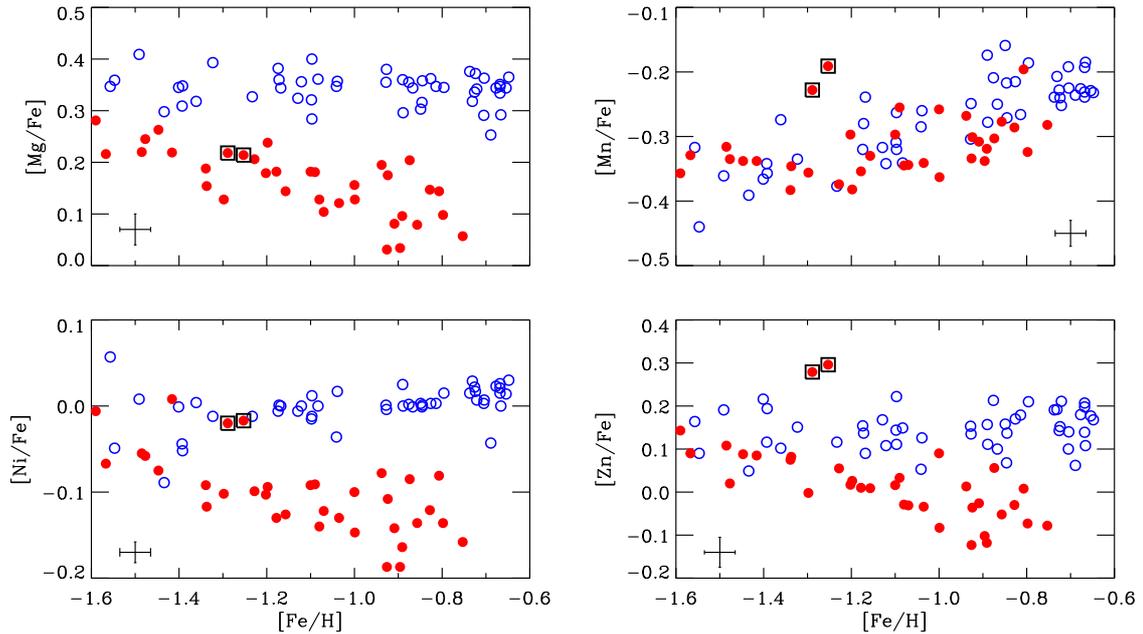}
\caption{\mgfe , \mnfe , \nife , and \znfe\ as a function of \feh\ for stars in
\citet{nissen10, nissen11}. Stars belonging to the
high-$\alpha$ sequence are shown with blue circles and those on the low-$\alpha$ sequence with
filled red circles. The components of G\,112-43/44 are indicated with a black square.
Typical 1\,$\sigma$ errors of the differential abundance ratios are shown at the bottom of the figures.}
\label{fig:Mg.Mn.Ni.Zn-feh}
\end{figure*}

We have also used $Gaia$ EDR3 proper motions and parallaxes to re-calculate space
velocities, $U_{\rm LSR}$, $V_{\rm LSR}$, $W_{\rm LSR}$, with respect to the
Local Standard of Rest (LSR) of the other stars in \citet{nissen10}.
This has led to a reduction of the errors of the space velocities from typically
10--20\,\kmprs\ to 1--2\,\kmprs\
mainly because the $Gaia$ distances are much more precise than
the $Hipparcos$ and photometric distances applied in the 2010 paper\,\footnote{Exceptions are
HD\,106516, HD\,163810, and HD\,219617 for which $Gaia$ parallaxes are not available.}.
Furthermore, we have updated the solar motion relative to the LSR to 
($U_{\sun}$, $V_{\sun}$, $W_{\sun}$) = (11.1, 12.24, 7.25)\,\kmprs\
\citep{schoenrich10} and the circular speed of the LSR to 232.8\,\kmprs
\citep{mcmillan17}. Based on these new kinematical data, Fig. \ref{fig:VphiVz} shows
the $V_{\phi}$--$V_{Z}$ diagram for stars in \citet{nissen10} that 
could be clearly classified as belonging
to either the high-$\alpha$ population (blue circles) or the low-$\alpha$ sequence (filled red circles).

\begin{table*}
\caption[ ]{Atmospheric parameters and abundance ratios for G\,112-43/44 and the
comparison stars}
\label{table:abundances}
\centering
\setlength{\tabcolsep}{0.07cm}
        \begin{tabular}{lccccccccccc}
        \hline\hline
        \noalign{\smallskip}
Atm.param. & G\,112  & G\,112  & mean of      &  CD$-$51& CD$-$61 & G\,005   &  G\,176 & G192  &   HD    & mean of & Diff.\tablefootmark{a} of   \\
/abun.rat.   & $-$43   & $-$44   & G\,112-43/44 &  4628 & 0282  & $-$19    & $-$53   & $-$43   &  163810 & comp. stars & $\langle \xfe \rangle$ \\
        \noalign{\smallskip}
        \hline
        \noalign{\smallskip}
        \teff [K] &   6074  &   5819  & 5947   &   6153  &   5759  &   5854  &   5523  &   6170  &   5501 & 5827  & ... \\
         \logg    &   4.03 &    4.25 &  4.14  &   4.31 &    4.31 &    4.26 &    4.48 &    4.29 &    4.56 &  4.37  & ... \\
     \turb [km/s] &   1.3  &    1.2  &  1.25  &   1.4  &   1.3   &    1.3  &   1.0   &    1.5  &    1.3  &  1.20  & ... \\        
         \feh     &  $-$1.253&   $-$1.289& $-$1.271  &  $-$1.298&   $-$1.228&   $-$1.178&   $-$1.337&   $-$1.339&   $-$1.202& $-$1.264  & ... \\
        \cfe\  & +0.018 & +0.050 & $+0.034 \pm 0.049$& $-$0.106 & $-$0.090 & $-$0.108 & ...    &  ...   & +0.005 & $-0.075 \pm 0.027$&  $+0.109 \pm 0.056$\\
        \ofe\  & +0.386 &  ...   & $+0.386 \pm 0.055$& +0.401 & +0.542 & +0.442 & +0.594 & ...    & +0.522 & $+0.500 \pm 0.035$&  $-0.114 \pm 0.065$\\
        \nafe\ & $-$0.108 & $-$0.117 & $-0.112 \pm 0.021$& $-$0.258 & $-$0.197 & $-$0.254 & $-$0.361 & $-$0.209 & $-$0.217 & $-0.249 \pm 0.024$&  $+0.137 \pm 0.032$\\
        \mgfe\ & +0.214 & +0.218 & $+0.216 \pm 0.021$& +0.128 & +0.206 & +0.182 & +0.154 & +0.188 & +0.179 & $+0.173 \pm 0.011$&  $+0.043 \pm 0.024$\\
        \sife\ & +0.152 & +0.203 & $+0.177 \pm 0.021$& +0.191 & +0.204 & +0.166 & +0.147 & +0.235 & +0.165 & $+0.185 \pm 0.013$&  $-0.007 \pm 0.025$\\
        \cafe\ & +0.291 & +0.259 & $+0.275 \pm 0.014$& +0.312 & +0.282 & +0.259 & +0.253 & +0.329 & +0.276 & $+0.285 \pm 0.012$&  $-0.010 \pm 0.019$\\
        \tife\ & +0.289 & +0.221 & $+0.255 \pm 0.014$& +0.244 & +0.170 & +0.135 & +0.152 & +0.292 & +0.225 & $+0.203 \pm 0.025$&  $+0.052 \pm 0.029$\\
        \crfe\ & $-$0.001 & $-$0.008 & $-0.005 \pm 0.014$& +0.004 & $-$0.023 & $-$0.049 & $-$0.013 & $-$0.047 & +0.024 & $-0.017 \pm 0.012$&  $+0.013 \pm 0.018$\\
        \mnfe\ & $-$0.191 & $-$0.228 & $-0.209 \pm 0.014$& $-$0.356 & $-$0.374 & $-$0.354 & $-$0.346 & $-$0.383 & $-$0.297 & $-0.352 \pm 0.012$&  $+0.142 \pm 0.019$\\
        \nife\ & $-$0.017 & $-$0.020 & $-0.019 \pm 0.008$& $-$0.102 & $-$0.099 & $-$0.130 & $-$0.117 & $-$0.092 & $-$0.103 & $-0.107 \pm 0.006$&  $+0.089 \pm 0.010$\\
        \cufe\ & $-$0.390 & $-$0.410 & $-0.400 \pm 0.025$& ...    & $-$0.630 & $-$0.580 & $-$0.590 &  ...   & $-$0.500 & $-0.575 \pm 0.029$&  $+0.175 \pm 0.038$\\
        \znfe\ & +0.296 & +0.279 & $+0.288 \pm 0.025$& $-$0.002 & +0.055 & +0.010 & +0.082 & +0.075 & +0.017 & $+0.040 \pm 0.015$&  $+0.248 \pm 0.029$\\
        \yfe\  & $-$0.139 & $-$0.250 & $-0.194 \pm 0.025$& $-$0.112 & $-$0.163 & $-$0.120 & $-$0.262 & $-$0.091 & $-$0.110 & $-0.143 \pm 0.026$&  $-0.051 \pm 0.036$\\
        \bafe\ & $-$0.273 & $-$0.288 & $-0.280 \pm 0.025$& $-$0.163 & $-$0.238 & $-$0.171 & $-$0.259 & $-$0.173 & $-$0.161 & $-0.194 \pm 0.017$&  $-0.086 \pm 0.030$\\
        \noalign{\smallskip}
        \hline
        \end{tabular}
        \tablefoot{
        \tablefoottext{a} {Difference between the mean \xfe\ for G\,112-43/44 and the comparison stars.}}
        \tablebib{\cfe\ and \ofe\ refer to 3D non-LTE values from \citet{amarsi19} and \cufe\ to 1D non-LTE values from \citet{yan16}. 
                  For all other abundance ratios, 1D LTE values from \citet{nissen10, nissen11} are given.}
        \end{table*}

As seen from Fig. \ref{fig:VphiVz}, G\,112-43/44 stands out from the other low-$\alpha$ stars by having 
a large (negative) velocity component perpendicular to the Galactic plane and by falling in
the $V_{\phi} - V_Z$ box of one of the streams
in the solar neighbourhood discovered by \citet{helmi99}. The two components
are in fact included among the 40 core members of the Helmi streams with distances less than 1\,kpc
listed in \citet[][Table 1]{koppelman19b} by their $Gaia$ DR2 numbers and were also
associated with the Helmi streams by \citet{jean-baptiste17} based on $Hipparcos$ data.
It is, therefore, likely that G\,112-43/44 was formed in the progenitor galaxy
of the Helmi streams. According to \citet{koppelman19b},  
this dwarf galaxy had a stellar mass of $\sim 10^8 \, M_{\sun}$ and was accreted 5--8\,Gyr ago.

\section{Element abundances}
\label{abundances}
The components of G\,112-43/44 were included among 94 halo and thick-disk stars
for which \citet{nissen10, nissen11} derived element abundances from 
high signal-to-noise, S/N\,$\sim 200 - 300$,
VLT/UVES and NOT/FIES spectra
using 1D MARCS model atmospheres \citep{gustafsson08}
to analyse the measured equivalent widths under the assumption of
Local Thermodynamic Equilibrium (LTE). As described in detail in \citet{nissen10},
the effective temperature, \teff , the surface gravity, \logg , and the microturbulence, \turb , 
of a given star, were determined from the condition that
the derived Fe abundances should have no systematic dependence on excitation potential, ionisation
stage, and  reduced equivalent width of the Fe lines. 
These parameters were determined with internal precisions of $\sigma \teff \simeq 30$\,K,
$\sigma \logg \simeq 0.05$\,dex, and $\sigma \turb \simeq  0.1$\,\kmprs and differential
abundance ratios, \xfe , for 12 elements (Na, Mg, Si, Ca, Ti, Cr, Mn, Ni, Cu, Zn, Y, and Ba) were
determined with 1\,$\sigma$ precisions of 0.01 to 0.04\,dex depending on the number 
of lines available for a given element \citep[see list of lines in Table 3 of][]{nissen11}.

G\,112-43/44 was found to belong to the low-$\alpha$ population 
but showed an overabundance of Mn/Fe, Ni/Fe, and Zn/Fe relative to the other low-$\alpha$ stars
(see Fig. \ref{fig:Mg.Mn.Ni.Zn-feh}). In the following we estimate the
significance of these enhancements and discuss if G\,112-43/44 deviates in other
abundance ratios relative to low-$\alpha$ stars with similar metallicity.

Table \ref{table:abundances} lists the atmospheric parameters and abundance ratios
determined for G\,112-43/44 and six low-$\alpha$ stars having \feh\ values within $\pm 0.10$ dex of
the mean \feh\ for G112-43/44. For most elements, the abundance ratios are LTE values from 
\citet{nissen10, nissen11},
but in the case of C and O, the \xfe\ values are from the 3D non-LTE analysis by \citet{amarsi19} of
two high-excitation \CI\ lines ($\lambda 5052.17$
and $\lambda 5380.14$\,\AA ) and the \OI\ triplet at $\lambda 7774$\,\AA ,
and in the case of Cu, abundances were adopted from \citet{yan16},
who made a 1D non-LTE analysis of \CuI\ lines in the spectra of \citet{nissen11}.
Actually, there is one additional low-$\alpha$ star, G\,053-41, having a
metallicity near that of  G\,112-43/44, but it has a very high Na/Fe ratio and low
C/Fe and O/Fe ratios suggesting that it was born as a second-generation star in
a globular cluster \citep{ramirez12, nissen14}. G\,053-41 is, therefore, excluded from
the comparison with G\,112-43/44. 

In Table \ref{table:abundances}, we compare the 
mean \xfe\ values for G\,112-43/44 (Col. 4) with the mean \xfe\ of the six comparison
stars (Col. 11). In the case of G\,112-43/44, the quoted errors of the mean \xfe\
are calculated as $\sigma \, \xfe / \sqrt{n}$, where $\sigma \, \xfe $ is the statistical
1\,$\sigma$ error estimated from the line-to-line scatter and $n = 2$ except in the case of
oxygen for which $n = 1$, because the O-triplet was not covered by the spectrum
of G\,112-44.  For the six comparison stars,
the quoted errors are the standard deviation of the mean of \xfe . Based on these data,
the last column gives the difference between the mean \xfe\ for  
G\,112-43/44 and the mean \xfe\ for the six comparison stars. The errors given are
calculated as the quadratic sum of the errors of the two mean values. 

The last column of Table \ref{table:abundances} shows
that the \mnfe , \nife , and \znfe\ values of G\,112-43/44 are enhanced 
relative to the corresponding mean values for the comparison stars at a 
confidence level of 7-8\,$\sigma$. \nafe\ and \cufe\
also seem significantly  enhanced (at a confidence level of 4-5\,$\sigma$). Furthermore,
\bafe\ may be lower in G\,112-43/44 than in the comparison stars, but this is 
only significant at a level of 3\,$\sigma$.  For the remaining
elements, the mean \xfe\ of  G\,112-43/44 agrees with the mean of \xfe\ for
the comparison stars within $\pm 2 \, \sigma$ of the estimated errors of the difference.

\begin{figure}
\centering
\includegraphics[width=8.0cm]{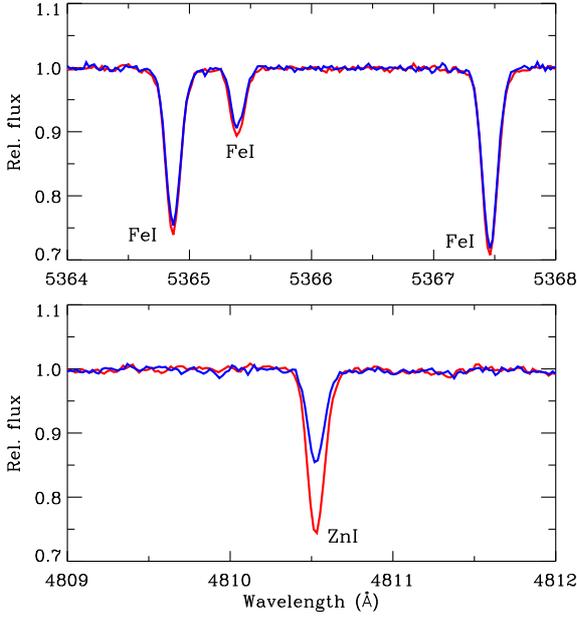}
\caption{Comparison of the spectrum of G\,112-43 (red line) with that of CD$-51 \,4628$
(blue line) for three FeI lines  around 5366\,\AA\ and the \ZnI\ line at 4810.5\,\AA .}
\label{fig:spectra}
\end{figure}

The largest enhancement occurs for Zn with $ \Delta \langle \xfe \rangle \, = \,0.248$.
This can be seen directly from the spectra as shown in Fig. \ref{fig:spectra},
where the spectrum of G\,112-43 is compared with that of one of the 
comparison stars, CD$-51 \,4628$. The atmospheric parameters of the two stars
are similar, i.e. ($\teff , \logg , \feh$) = (6074\,K, 4.03, $-1.253$) for G\,112-43
and (6153\,K, 4.31, $-1.298$) for CD$-51 \,4628$. The differences in these parameters have
only a minor effect on the strengths of \FeI\ and \ZnI\ lines. As seen, the \FeI\
lines are only slightly stronger in the spectrum of G\,112-43, whereas the 
equivalent width of the \ZnI\ line is nearly a factor of two larger.  

As mentioned above, the abundances in \citet{nissen10, nissen11} were derived by 
assuming LTE. 
Deviations from LTE (non-LTE effects) may affect the derived 
abundances and hence the trends of \xfe\ a function of \feh , but the 
effects on differential abundances at a given \feh\ were expected to be small,
because the stars are confined to small ranges in \teff\ and \logg.
Thanks to recent detailed non-LTE studies,
we now have the possibility to check this expectation in connection
with the determination of abundances of G\,112-43/44 relative to the comparison stars.  

Table \ref{table:non-LTE corr} lists the non-LTE corrections relative to the solar corrections, i.e. 
$\delta \xh \, = \, \xh_{\rm non-LTE} - \xh_{\rm LTE}$, for G\,112-43 and G\,112-44
in Cols. 3 and 4 and the mean correction for the comparison stars 
in Col. 5. The last column gives the difference of the mean correction 
of \xfe\ between G\,112-43/44 and the comparison stars.
These corrections were obtained for the applied spectral lines
by interpolation in tables of non-LTE corrections
in the listed references\,\footnote{For the Bergemann et al. 
references, {\tt Spectrum Tools} at {\tt http://nlte.mpia.de} was used for the interpolation.}.

As seen from Table \ref{table:non-LTE corr}, 
the differences of the non-LTE corrections between G\,112-43 and G\,112-44 are quite small,
i.e. $\sim -0.01$ to $\sim +0.03$\,dex, except in the case of Cu for which the difference
reaches +0.06\,dex, which is caused by steeply rising non-LTE corrections for
the \CuI\ lines as a function of increasing \teff\ and  decreasing
\logg\,\footnote{G\,112-43 is a more evolved star with 
$\delta \teff \, = \,255$\,K and $\delta \logg \, = \, -0.22$\,dex relative to G\,112-44.}.
For oxygen there is a very significant non-LTE effect on the difference of \xfe\
between  G112-43 and the comparison stars (last column),
but for the other elements the non-LTE effects on the differences in \xfe\ between
G\,112-43/44 and the comparison stars lie within the
statistical uncertainties given in the last column of 
Table \ref{table:abundances}. In particular, we note that non-LTE effects have a
negligible influence on the derived overabundance of Mn/Fe, Cu/Fe, and Zn/Fe in
G\,112-43/44 relative to the comparison stars. Unfortunately, there is no non-LTE 
study of Ni (and Y), but it would be a surprise if the measured overabundance of 
Ni/Fe in the G\,112-43/44 pair has anything to do with non-LTE effects.
The good agreement between the Ni abundance in the solar photosphere derived from an
LTE analysis of \NiI\ lines \citep{scott15} and the
meteoritic abundance \citep{lodders03} suggests that non-LTE effects on \NiI\ lines are small.
Concerning yttrium, the abundances were derived from two lines belonging to 
the \YII\ majority species, and we would therefore expect that the non-LTE effects on the
derived differential Y abundances are small like in the case of Ba abundances derived from   
\BaII\ lines.
  
A notorious uncertainty in non-LTE calculations is the rate of collisions of atoms
with neutral hydrogen. For some of the elements in Table \ref{table:non-LTE corr} 
(C, O, Mg, Ca, and Fe),
the non-LTE studies are based on new quantum-mechanical rate coefficients, but for the other
elements, the classical coefficients of \citet{drawin69} (eventually scaled by an empirical
factor) were applied, which makes the non-LTE corrections somewhat uncertain. Furthermore, 
the corrections were calculated for 1D atmospheric models except in the case of C and O, for which
\citet{amarsi19} have provided 3D\,non-LTE -- 1D\,LTE corrections. 
Hence, there is room for improvements, but we do not expect that 3D--1D corrections
will have a significant effect on the differential abundances of G\,112-43/44 relative
to the comparison stars given that the stars have nearly the same \feh\ and have only small
differences in \teff\ and \logg .

\begin{table}
\caption[ ]{Non-LTE corrections of abundances}
\label{table:non-LTE corr}
\centering
\setlength{\tabcolsep}{0.15cm}
\begin{tabular}{lccccc}
\hline\hline
\noalign{\smallskip}
 Elem.    & Ref.& G112-43      & G112-44      & Comp.stars    &  Difference \tablefootmark{a}  \\
     &     & $\delta \xh$ & $\delta \xh$ & $\langle \delta \xh \rangle$  &  $\langle \delta \xfe \rangle$   \\
\noalign{\smallskip}
\hline
 \FeI & 1 &+0.109 &+0.088 & +0.082 & ...       \\
 \CI  & 2 &$-$0.035 &$-$0.037 & $-$0.035 & $-$0.017  \\
 \OI  & 2 &$-$0.012 &  ...    & +0.074  & $-$0.113  \\
 \NaI & 3 &+0.016 &+0.024 & +0.029 & $-$0.025  \\
 \MgI & 4 &+0.040 &+0.017 & +0.014 & $-$0.002  \\
 \SiI & 5 &$-$0.004 &+0.000 & +0.000 & $-$0.018  \\
 \CaI & 6 &+0.062 &+0.048 & +0.042 & $-$0.003  \\
 \TiI & 7 &+0.216 &+0.192 & +0.176 & +0.012  \\
 \CrI & 8 &+0.179 &+0.162 & +0.152 & +0.002  \\
 \MnI & 9 &+0.230 &+0.226 & +0.214 & $-$0.002  \\
 \CuI & 10 &+0.160&+0.100 & +0.100 & +0.014   \\
 \ZnI & 11 &+0.125 &+0.096 & +0.087 & +0.007  \\
 \BaII& 12 &+0.042 &+0.024 & +0.024 & $-$0.007 \\
\noalign{\smallskip}
\hline
\end{tabular}
\tablefoot{
\tablefoottext{a} {Difference between the mean non-LTE correction of \xfe\ for G\,112-43/44 
and the comparison stars. While Cols. 3, 4, and 5 refer to non-LTE corrections of \xh ,
Col. 6 gives the corrections of \xfe .}
}
\tablebib{
(1)~\citet{amarsi16b}; (2)~\citet{amarsi19}; (3)~\citet{lind11}, (4)~\citet{bergemann17}; (5)~\citet{bergemann13};
(6)~\citet{mashonkina17}; (7)~\citet{bergemann11}; (8)~\citet{bergemann10}; 
(9)~\citet{bergemann08}; (10)~\citet{yan16}; (11)~\citet{takeda05b}; (12)~\citet{korotin15}.
}
\end{table}

\begin{figure}
\centering
\includegraphics[width=8.0cm]{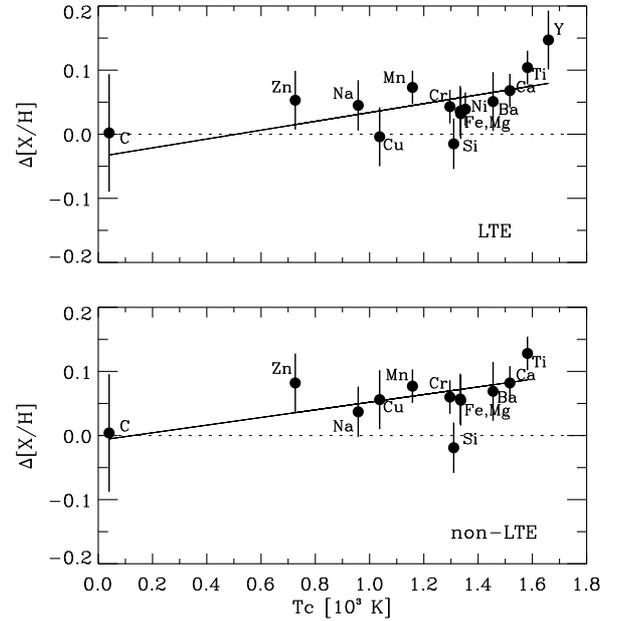}
\caption{Difference of \xh\ between G\,112-43 and G\,112-44 as a function of
elemental condensation temperature. The upper panel refers to LTE abundances, whereas non-LTE
corrections from Table \ref{table:non-LTE corr} have been applied in the lower panel. The solid
lines correspond to the fits of the data given in Eqs. (1) and (2).}
\label{fig:xh-Tc}
\end{figure}

\begin{figure*}
\centering
\includegraphics[width=17cm]{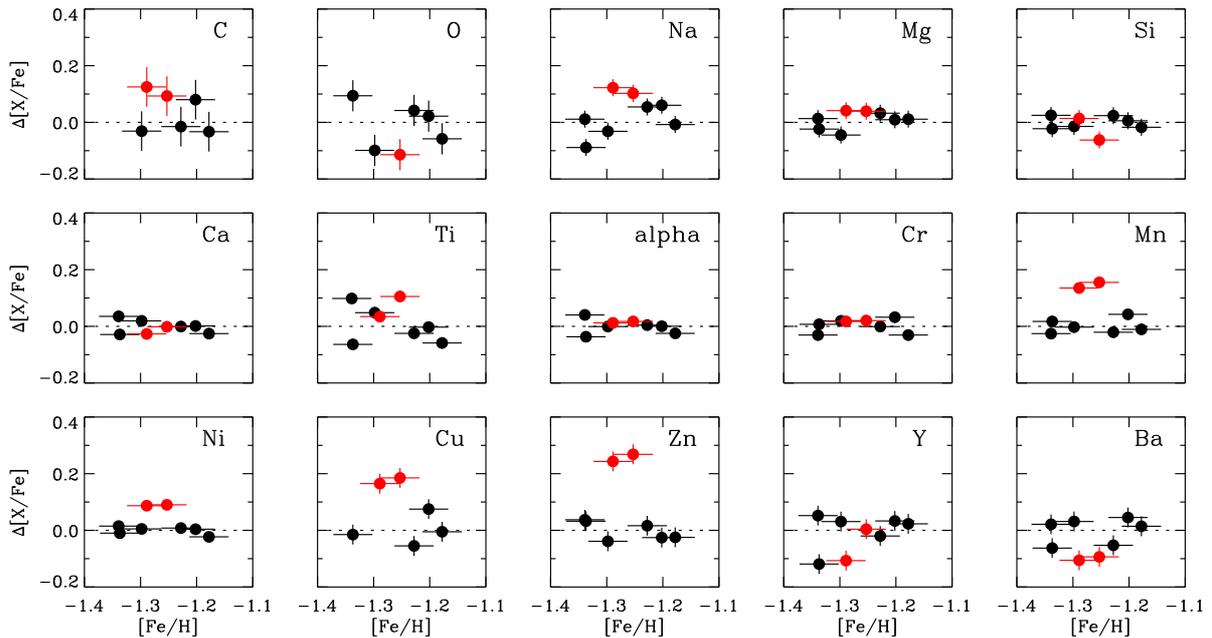}
\caption{Difference in \xfe\ for stars in Table \ref{table:abundances}
relative to the mean of \xfe\ for the six comparison stars as a function of  \feh .
G\,112-43 and G\,112-44 are shown with red filled circles and the comparison stars
with black filled circles. The error bars refer to the 1\,$\sigma$ uncertainties of the abundance ratios.
Non-LTE corrections of \xfe\ are included when available.}
\label{fig:G112-comparison}
\end{figure*}

\section{Discussion}
\label{discussion}
\subsection{Abundance differences between G\,112-43 and G\,112-44}
\label{comparison1}
As seen from Table \ref{table:abundances}, the abundances of the two components of 
G\,112-43/44 show good agreement. For most elements, the difference is less than, or in the
order of, the 1\,$\sigma$ error of the abundance ratios. As shown in Fig. \ref{fig:xh-Tc},
the difference in \xh\ between G\,112-43 and G\,112-44 seems, however, to depend on
elemental condensation temperature \Tc\ \citep{lodders03}. Based on a weighted least squares fit
to the LTE abundances, we find a linear relation
\begin{eqnarray}
\Delta \xh = -0.035 + 6.90 \, (\pm 3.67) 10^{-5} \cdot  \Tc \, {\rm dex\,K}^{-1}.
\end{eqnarray}
If the non-LTE corrections in Table \ref{table:non-LTE corr} are applied the relation becomes
\begin{eqnarray}
\Delta \xh = -0.008  + 5.96 \, (\pm 3.84) 10^{-5} \cdot  \Tc \, {\rm dex\,K}^{-1}.
\end{eqnarray}
In this case, Ni and Y were excluded from the fit, because the
non-LTE corrections are not available for these elements. 
As seen, the non-LTE corrections have only a
small effect on the $\Delta \xh -\Tc$ relation. The derived 
slope of $\Delta \xh$ as a function of \Tc\ is, however, only significant at
the $1.9 \, \sigma$ level in the LTE case and $1.6 \, \sigma$ in the non-LTE case.

Significant \Tc\ trends of the difference in \xh\ between components of 
comoving twin-stars have been discovered in at least seven cases
as reviewed by \citet{ramirez19}, but all these stars belong to the disk
population and have metallicities ranging from $\feh = -0.4$ to 0.4. 
For the halo population, there is only one co-moving pair, HD\,134439/134440 with $\feh \simeq -1.4$,
for which a \Tc\ trend of chemical abundances has been claimed 
\citep{chenyuqin06, chenyu14}, but \citet{reggiani18} find no trend of 
$\Delta \xh$ with \Tc\ based on a high-precision differential abundance analysis of
high-resolution ($R \simeq 72\,000$) HDS/Subaro spectra with S/N$\sim 250$.
They found, however, an average difference in \xh\ of 0.06\,dex
between HD\,134440 and HD\,134439 for 17 elements ranging
from C to Ba, which they suggest could be due to engulfment of a Jupiter-mass planet
by HD\,134440. In the case of G\,112-43/44, the $\Delta \xh$-\Tc\ trend 
in Fig. \ref{fig:xh-Tc} could be due to sequestration of refractory elements
\citep{melendez09} in planets around  G\,112-44 or to accretion of 
earth-like material \citep{cowley21} into the convection zone of G\,112-43.
Alternative explanations include dust-gas separation in star-forming clouds
\citep{gustafsson18} or in proto-planetary disks \citep{booth20}.

\subsection{The enhancement of Mn, Ni, Cu, and Zn abundances in G\,112-43/44}
\label{comparison2}
The \xfe\ abundances ratios for G\,112-43/44 and the comparison stars are shown
in Fig. \ref{fig:G112-comparison} on a scale where the mean
\xfe\ of the comparison stars is normalised to zero for all elements.
The error bars refer to the statistical uncertainties of the differential abundance ratios
as estimated in \citet{nissen10, nissen11} and \citet{amarsi19}.
These errors vary from element to element depending on the number and strengths of lines
available for the abundance determination. The errors are, for example, very small
for nickel, for which about 25 \NiI\ lines are available, but are relatively
large for carbon because only two
weak \CI\ lines could be applied. The errors are also large for the oxygen abundances,
because the \OI\ 7774\,\AA\ triplet lines are sensitive to errors in
\teff\ and \logg .

Before making Fig. \ref{fig:G112-comparison}, we have included non-LTE
corrections of \xfe\ (except for Ni and Y) for all stars, but this has only a small
effect on the mean difference in \xfe\ between G\,112-43/44 and the comparison stars
(see last column of Table \ref{table:non-LTE corr}).
As seen from Fig. \ref{fig:G112-comparison}, the overabundances of 
\mnfe , \nife , \cufe , and \znfe\ in G\,112-43/44 look very significant,  whereas the possible
deviations of \nafe\ and \bafe\ are less convincing,
because of a relatively large scatter in these abundance ratios among the comparison stars.
Furthermore, we note the
excellent agreement between \alphafe\ for the  G\,112-43/44 pair and the comparison stars. 

G\,112-43 (but not G\,112-44)  and one of our comparison stars (G\,176-53) were included
in a high-precision study of chemical abundances in the Milky Way thick disk and halo by  
\citet{ishigaki12, ishigaki13}. As discussed in their papers, the derived \xfe\ values show
good agreement with the corresponding values in \citet{nissen10, nissen11}. In particular, 
Ishigaki et al. found the same high Zn/Fe ratio of $\sim 0.3$\,dex for G\,112-43 as in the present paper,
and they also found \mnfe\ and \nife\ to be enhanced in G\,112-43 relative to  G\,176-53
(the abundance of Cu in G\,176-53 was not determined by Ishigaki et al.).

Interestingly, three other $\alpha$-poor stars
in the Galactic halo have the same pattern of Mn, Ni, and Zn  
enhancements relative to Fe as G\,112-43/44. \citet{ivans03} found G\,4-36
(a turn-off star with $\teff = 5975$\,K and $\feh = -1.94$) and BPS\,CS\,22966-43,
(a blue metal-poor star with $\teff = 7400$\,K and $\feh = -1.91$) 
to be enhanced in Mn/Fe, Ni/Fe, and Zn/Fe relative to Milky Way stars with similar metallicity
(see their Figs. 10 and 11), and \citet{honda11} found BPS\,BS\,16920-17 (a giant star
with $\teff = 4760$\,K and $\feh = -3.1$) to have similar enhancements relative to a
well-known halo giant, HD\,4306, having $\teff = 4810$\,K and $\feh = -2.8$.
For all three stars, the enhancement in Zn is $\Delta \znfe \sim 0.9$
and $\Delta \mnfe$ and  $\Delta \nife$ are at a level of 0.5\,dex. 
The abundance of Cu was not determined for the three stars.

High Zn/Fe ratios, $\znfe \sim 0.5$, in very metal-poor stars \citep{cayrel04, nissen07} have been
explained as due to hypernovae \citep{umeda02}, that is massive core-collapse supernovae with
explosion energies $E \sim 10^{52}$\,erg, ten times higher than normal Type II SNe.
However, a hypernova cannot explain the enhancement of Mn/Fe, Ni/Fe, and Cu/Fe;
as seen from Fig. 8 in the review of nucleosynthesis by \citet{nomoto13}, the yields
of \mnfe\ and \cufe\ are around $-0.7$\,dex and that of \nife\ is $\sim \, -0.2$\,dex.
Figure 8 in \citet{nomoto13} also shows that pair-instability SNe, faint SNe, and the 
classical W7 single-degenerate Chandrasekhar-mass ($M_{\rm Ch}$) Type Ia SN model of \citet{iwamoto99} all fail
to explain the distribution of $\Delta \xfe$ for  Mn, Ni, Cu, and Zn in G\,112-43/44
and the three $\alpha$-poor stars discussed above. There are, however, other models
of Type Ia SNe with different nucleosynthesis yields as recently analysed by
\citet{lach20}. Among the various possible scenarios including single and double 
degenerate models with both $M_{\rm Ch}$  and sub-$M_{\rm Ch}$ masses and different explosion mechanisms,
for which \xfe\ is shown in their Fig. 3, one model produces enhanced \xfe\ values for
Mn, Ni, Cu, and Zn. This so-called HeD-S model consists of a sub-$M_{\rm Ch}$ mass
white dwarf with a prominent helium shell in which a detonation takes place without triggering
explosion in the CO core.

\begin{figure}
\centering
\includegraphics[width=8.0cm]{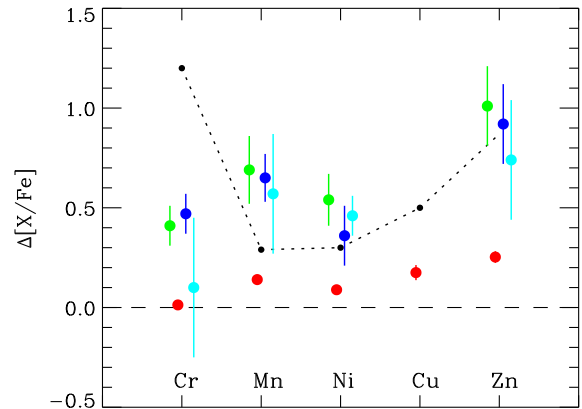}
\caption{Enhancements of iron-peak elements for G\,112-43/44 
and three $\alpha$-poor halo stars in comparison
with the \xfe\ yield distribution (dotted line) calculated by \citet{lach20}
for a pure helium shell detonation Type Ia supernova model.   
Red symbols refer to data for G\,112-43/44, blue symbols to
G\,4-36, green symbols to  BPS\,CS\,22966-43,
and cyan symbols to BPS\,BS\,16920-17. The error bars refer to statistical
uncertainties in the abundance determinations.}
\label{fig:CrMnNiCuZn}
\end{figure}

The calculated \xfe\ yields of iron-peak elements\,\footnote{Cobalt is not included, because its
abundance has not been determined for G\,112-43/44 and BPS\,CS\,22966-43 and is very uncertain for G\,4-36 and 
BPS\,BS\,16920-17} for the HeD-S model is shown in 
Fig. \ref{fig:CrMnNiCuZn} in comparison with the observed $\Delta \xfe$ values 
for G\,112-43/44 and the three $\alpha$-poor halo stars. 
As seen, there is satisfactory agreement between the HeD-S model predictions
and the observed enhancements of \nife\ and \znfe\ for the three $\alpha$-poor stars,
whereas the observed enhancement of Mn/Fe is a bit higher.
The \xfe\ enhancements for G\,112-43/44 (including \cufe ) are not as large as the 
\xfe\ yield predictions for the HeD-S model but have a similar relative distribution.
Such an offset may occur if
the elements produced by the HeD-S supernova are mixed with a relatively
large amount of gas produced by the kind of `normal' Type Ia SNe controlling the
chemical evolution of iron-peak elements for the stellar system in which G\,112-43/44
was born.

The HeD-S model cannot, however, explain all abundances in
G\,112-43/44 and the three $\alpha$-poor halo stars. As seen from Fig. \ref{fig:CrMnNiCuZn},
the predicted yield of \crfe\ is much higher than
the observed $\Delta \crfe$ enhancements, and the HeD-S models also predicts super-solar values of
\cafe\ and \tife , 0.5 and 1.8\,dex respectively, in strong disagreement with the much
lower observational values. It remains to be seen if this problem can be solved
with a revised model for a He shell detonation Type Ia supernova.
In any case, we think it is likely that the Mn/Fe, Ni/Fe, Cu/Fe, and Zn/Fe
enhancements in G\,112-43/44 and the three $\alpha$-poor 
stars are caused by element contribution from a special kind of 
supernovae to pockets of gas from which these stars
were formed. Such inhomogeneous mixing of supernovae products is
more likely to affect the chemical evolution of dwarf galaxies than higher-mass systems,
because in lower-mass galaxies, the gas has a relatively long cooling time resulting in 
episodic star-formation bursts \citep{revaz09, venn12}. 
In view of this, it is interesting that G\,112-43/44 as a member of the Helmi streams
probably was born in a dwarf galaxy and 
that one of the three $\alpha$-poor halo stars with enhancements of Mn/Fe, Ni/Fe, and
Zn/Fe also have Helmi streams kinematics\,\footnote{Space velocities were calculated
based on $Gaia$ EDR3 data supplemented with radial velocities,
RV = $-277 \pm 10$\,\kmprs\ for BPS\,CS\,22966-43 \citep{wilhelm99} and RV = $-210 \pm 10$\,\kmprs\
for BPS\,BS\,16920-17 \citep{allende.prieto00}.} as seen from Fig. \ref{fig:VphiVz}.
G\,4-36 and  BPS\,BS\,16920-17 have kinematics like members of the Gaia-Enceladus population, but
BPS\,CS\,22966-43 lies in the Helmi box with positive $V_{Z}$. 

Not all stars belonging to the Helmi streams have unusual abundances.  
\citet{roederer10} made a detailed abundance analysis of 12
subgiant and giant stars with Helmi streams kinematics and \citet{gull21} 
recently added seven stars to this work.
Within the typical errors of the abundance determinations (0.10 to 0.15\,dex), \mnfe , \nife , \cufe ,
and \znfe , agree with the \xfe -\feh\ trends of Galactic halo stars. 
The stars investigated have, however, $\feh \simlt -1.5$ and high \alphafe\
ratios, e.g. $\mgfe \sim 0.3$, so it remains to be seen if the Helmi streams contain
a population of low-$\alpha$ stars at higher metallicities and if some of these 
stars have overabundances of Mn, Ni, Cu, and Zn relative to Fe. 

Low-$\alpha$ stars are well known in dwarf galaxies as reviewed by \citet{tolstoy09}. 
In the Sculptor dSph galaxy, for example, \mgfe\ shows a declining
metallicity trend from $\mgfe \simeq 0.4$ at $\feh = -1.8$ to  $\mgfe \simeq -0.2$ 
at $\feh = -1.0$. Interestingly, \citet{skuladottir17} find indications of a cosmic
scatter in \znfe\ of the Sculptor stars, i.e. from about $-0.6$\,dex to +0.4\,dex and
a positive correlation between \nife\ and \znfe .  At a distance of $\sim 85$\,kpc, the giants
observed in Sculptor are, however, faint ($17.0 < V < 18.5$), and the errors of 
\znfe\ are in the order of $\pm 0.3$\,dex. It will require more precise 
\nife\ and \znfe\ values and also improved  precision of \mnfe\ 
\citep{north12} to verify if stars with enhanced
Mn/Fe, Ni/Fe, and Zn/Fe ratios are present in Sculptor.
 
Another interesting case is the ultrafaint dwarf galaxy Horologium I for which
\citet{nagasawa18} found three very metal-poor ($\feh \sim -2.5$) giant stars to be $\alpha$-poor 
(\mgfe\ and \cafe\ $\simeq 0.0$) and  enhanced in \mnfe\ by about 0.4 dex relative to 
Galactic halo stars. \nife\ seems, however, to be normal, and Cu and Zn abundances were not determined.
Again, we need more precise abundances including those of Cu and Zn to make conclusions about
the possible existence of stars in Horologium I  with peculiar abundance ratios among
the iron-peak elements.

\section{Summary and conclusions}
\label{conclusions}
In this paper, we have made a high-precision study of elemental abundance ratios for
the components of the low-$\alpha$ metal-poor binary star G\,112-43/44 in comparison with  
abundance ratios for six low-$\alpha$ halo stars having nearly the same metallicity and similar
\teff\ and \logg\ values. Non-LTE effects on the derived differential abundance ratios were
considered, but found to be small, except for [O/Fe] (see Table \ref{table:non-LTE corr}).
As a main result, we find that
the abundance ratios of Mn, Ni, Cu, and Zn with respect to Fe are significantly 
enhanced in G\,112-43/44 relative to the comparison stars, i.e. $\Delta \mnfe = 0.14 \pm 0.02$, 
$\Delta \nife =0.09 \pm 0.01$, $\Delta \cufe = 0.18 \pm 0.04 $, 
and $\Delta \znfe = 0.25 \pm 0.03$. 

From a literature search, we found three 
other $\alpha$-poor halo stars, G\,4-36, BPS\,CS\,22966-43 \citep{ivans03}, and BPS\,BS\,16920-17
\citep{honda11} with enhanced abundances of Mn, Ni, and Zn relative to Fe. 
The errors of the abundance ratios are much higher than in our study of G\,112-43/44,
0.10 to 0.20\,dex, but the amplitude of the enhancements is also higher reaching
$\Delta \znfe \sim 1.0$, so the significance of the abundance peculiarities is still high.

Interestingly, two of the four stars with enhanced \xfe\ values for the iron-peak
elements (G\,112-43/44 and BPS\,CS\,22966-43) have 
very high velocity components perpendicular to the Galactic plane indicating that they
are members of the Helmi streams. This suggests that the occurrence of such stars
has a higher frequency in the progenitor dwarf galaxy of the Helmi streams than in the more 
massive Gaia-Enceladus galaxy responsible for most of the low-$\alpha$ halo stars in the solar
neighbourhood.
In view of this, it would be interesting to carry out a high-precision abundance 
study of K giants in dwarf galaxies with a large population of 
low-alpha stars such as Fornax and Sculptor. Due to the large distance of these
systems it will, however, require extremely large telescopes to obtain high-resolution
spectra with sufficiently high S/N to detect overbundances of
Mn, Ni, Cu, and Zn with respect to Fe. It would also be interesting to extend the 
abundance study by \citet{roederer10} of members of the Helmi streams to stars with  
$\feh > -1.5$ at which metallicy one would expect that Type Ia SNe start contributing to
the abundance of the iron-peak elements in the progenitor galaxy of the streams. 
We note, in this connection, that the metallicity distribution of 
members of the Helmi streams peaks at $\feh \simeq -1.5$ with
a tail up to $\feh \sim -0.5$ \citep{koppelman19b}.

It is unclear which SNe type can produce an overabundance of Mn, Ni, Cu, and Zn
relative to Fe. Hypernovae have been proposed \citep{umeda02} as the source of
enhanced Zn/Fe values in very metal-poor stars, but they cannot explain the
enhancement of Mn/Fe, Ni/Fe, and Cu/Fe \citep{nomoto13}. The pure helium shell
detonation Type Ia SN model of \citet{lach20} is a more promising candidate 
(see Fig. \ref{fig:CrMnNiCuZn}) but this model also predicts super-solar 
ratios of Ca/Fe, Ti/Fe, and Cr/Fe, which are not found for the four stars discussed
in this paper. It would be interesting to investigate if the helium shell detonation model
can be modified to produce solar ratios of Ca/Fe, Ti/Fe, and Cr/Fe while still producing 
enhanced ratios of Mn/Fe, Ni/Fe, Cu/Fe, and Zn/Fe.

As an additional result, we have found that the components of the  
G\,112-43/44 binary star do not have exactly the same abundance of elements and that
the difference seems to be correlated with elemental condensation temperature
(see Fig. \ref{fig:xh-Tc}). The $\Delta \xh$-\Tc\ slope is, however, only significant at
a confidence level of 1.6\,$\sigma$ when differential non-LTE corrections are applied.
One would need to increase the  S/N
of the spectra from $\sim 300$  in this paper to $\sim 600$ to decide if
the \Tc\ slope is significant. It would also be important to observe  
the \OI\ triplet at 7774\,\AA\ for both stars in order to determine
the difference in the O abundance, because oxygen has a low condensation temperature (\Tc\ = 180\, K)
like carbon (\Tc\ = 40\, K). If the $\Delta \xh$ - \Tc\ trend is confirmed, it would 
show that abundance differences correlated with \Tc\ also occur between components of wide-orbit binaries
belonging to the metal-poor halo population; this has already been found for several
binaries belonging to the thin disk population \citep{ramirez19}. Given that these trends may be
due to star-planet interactions, the detection of a $\Delta \xh$-\Tc\ trend 
for a halo binary star would be interesting.

\begin{acknowledgements}
The referee is thanked for a constructive and helpful report.
Funding for the Stellar Astrophysics Centre is provided by The
Danish National Research Foundation (Grant agreement no.: DNRF106).
This research has made use of the SIMBAD database,
operated at CDS, Strasbourg, France, as well as
data from the European Space Agency (ESA) mission
{\it Gaia} (\url{https://www.cosmos.esa.int/gaia}), processed by the {\it Gaia}
Data Processing and Analysis Consortium (DPAC,
\url{https://www.cosmos.esa.int/web/gaia/dpac/consortium}). Funding for the DPAC
has been provided by national institutions, in particular the institutions
participating in the {\it Gaia} Multilateral Agreement.

\end{acknowledgements}

\bibliographystyle{aa}
\bibliography{nissen.2021}

\begin{thebibliography}{81}
\expandafter\ifx\csname natexlab\endcsname\relax\def\natexlab#1{#1}\fi

\bibitem[{{Allende Prieto} {et~al.}(2000){Allende Prieto}, {Rebolo},
  {Garc{\'\i}a L{\'o}pez}, {Serra-Ricart}, {Beers}, {Rossi}, {Bonifacio}, \&
  {Molaro}}]{allende.prieto00}
{Allende Prieto}, C., {Rebolo}, R., {Garc{\'\i}a L{\'o}pez}, R.~J., {et~al.}
  2000, \aj, 120, 1516

\bibitem[{{Amarsi} {et~al.}(2016){Amarsi}, {Lind}, {Asplund}, {Barklem}, \&
  {Collet}}]{amarsi16b}
{Amarsi}, A.~M., {Lind}, K., {Asplund}, M., {Barklem}, P.~S., \& {Collet}, R.
  2016, \mnras, 463, 1518

\bibitem[{{Amarsi} {et~al.}(2019){Amarsi}, {Nissen}, \&
  {Sk{\'u}lad{\'o}ttir}}]{amarsi19}
{Amarsi}, A.~M., {Nissen}, P.~E., \& {Sk{\'u}lad{\'o}ttir}, {\'A}. 2019, \aap,
  630, A104

\bibitem[{{Bergemann}(2011)}]{bergemann11}
{Bergemann}, M. 2011, \mnras, 413, 2184

\bibitem[{{Bergemann} \& {Cescutti}(2010)}]{bergemann10}
{Bergemann}, M. \& {Cescutti}, G. 2010, \aap, 522, A9

\bibitem[{{Bergemann} {et~al.}(2017){Bergemann}, {Collet}, {Amarsi}, {Kovalev},
  {Ruchti}, \& {Magic}}]{bergemann17}
{Bergemann}, M., {Collet}, R., {Amarsi}, A.~M., {et~al.} 2017, \apj, 847, 15

\bibitem[{{Bergemann} \& {Gehren}(2008)}]{bergemann08}
{Bergemann}, M. \& {Gehren}, T. 2008, \aap, 492, 823

\bibitem[{{Bergemann} {et~al.}(2013){Bergemann}, {Kudritzki}, {W{\"u}rl},
  {Plez}, {Davies}, \& {Gazak}}]{bergemann13}
{Bergemann}, M., {Kudritzki}, R.-P., {W{\"u}rl}, M., {et~al.} 2013, \apj, 764,
  115

\bibitem[{{Booth} \& {Owen}(2020)}]{booth20}
{Booth}, R.~A. \& {Owen}, J.~E. 2020, \mnras, 493, 5079

\bibitem[{{Cayrel} {et~al.}(2004){Cayrel}, {Depagne}, {Spite}, {Hill}, {Spite},
  {Fran{\c{c}}ois}, {Plez}, {Beers}, {Primas}, {Andersen}, {Barbuy},
  {Bonifacio}, {Molaro}, \& {Nordstr{\"o}m}}]{cayrel04}
{Cayrel}, R., {Depagne}, E., {Spite}, M., {et~al.} 2004, \aap, 416, 1117

\bibitem[{{Chen} {et~al.}(2014){Chen}, {King}, \& {Boesgaard}}]{chenyu14}
{Chen}, Y., {King}, J.~R., \& {Boesgaard}, A.~M. 2014, \pasp, 126, 1010

\bibitem[{{Chen} \& {Zhao}(2006)}]{chenyuqin06}
{Chen}, Y.~Q. \& {Zhao}, G. 2006, \mnras, 370, 2091

\bibitem[{{Cowley} {et~al.}(2021){Cowley}, {Bord}, \& {Y{\"u}ce}}]{cowley21}
{Cowley}, C.~R., {Bord}, D.~J., \& {Y{\"u}ce}, K. 2021, \aj, 161, 142

\bibitem[{{Drawin}(1969)}]{drawin69}
{Drawin}, H.~W. 1969, Zeitschrift fur Physik, 225, 483

\bibitem[{{Fulbright}(2002)}]{fulbright02}
{Fulbright}, J.~P. 2002, \aj, 123, 404

\bibitem[{{Gaia Collaboration} {et~al.}(2018){Gaia Collaboration}, {Babusiaux},
  {van Leeuwen}, {Barstow}, {Jordi}, {Vallenari}, {Bossini}, {Bressan},
  {Cantat-Gaudin}, {van Leeuwen}, {Brown}, {Prusti}, {de Bruijne},
  {Bailer-Jones}, \& et.al}]{gaia.collaboration18a}
{Gaia Collaboration}, {Babusiaux}, C., {van Leeuwen}, F., {et~al.} 2018, \aap,
  616, A10

\bibitem[{{Gaia Collaboration} {et~al.}(2020){Gaia Collaboration}, {Brown},
  {Vallenari}, {Prusti}, {de Bruijne}, {Babusiaux}, \&
  {Biermann}}]{gaia.collaboration20}
{Gaia Collaboration}, {Brown}, A.~G.~A., {Vallenari}, A., {et~al.} 2020, arXiv
  e-prints, arXiv:2012.01533

\bibitem[{{Gaia Collaboration} {et~al.}(2016){Gaia Collaboration}, {Prusti},
  {de Bruijne}, {Brown}, {Vallenari}, {Babusiaux}, {Bailer-Jones}, {Bastian},
  {Biermann}, {Evans}, \& et~al.}]{gaia.collaboration16}
{Gaia Collaboration}, {Prusti}, T., {de Bruijne}, J.~H.~J., {et~al.} 2016,
  \aap, 595, A1

\bibitem[{{Gratton} {et~al.}(2003){Gratton}, {Carretta}, {Desidera},
  {Lucatello}, {Mazzei}, \& {Barbieri}}]{gratton03}
{Gratton}, R.~G., {Carretta}, E., {Desidera}, S., {et~al.} 2003, \aap, 406, 131

\bibitem[{{Gull} {et~al.}(2021){Gull}, {Frebel}, {Hinojosa}, {Roederer}, {Ji},
  \& {Brauer}}]{gull21}
{Gull}, M., {Frebel}, A., {Hinojosa}, K., {et~al.} 2021, arXiv e-prints,
  arXiv:2102.00066

\bibitem[{{Gustafsson}(2018)}]{gustafsson18}
{Gustafsson}, B. 2018, \aap, 620, A53

\bibitem[{{Gustafsson} {et~al.}(2008){Gustafsson}, {Edvardsson}, {Eriksson},
  {J{\o}rgensen}, {Nordlund}, \& {Plez}}]{gustafsson08}
{Gustafsson}, B., {Edvardsson}, B., {Eriksson}, K., {et~al.} 2008, \aap, 486,
  951

\bibitem[{{Hanson} {et~al.}(1998){Hanson}, {Sneden}, {Kraft}, \&
  {Fulbright}}]{hanson98}
{Hanson}, R.~B., {Sneden}, C., {Kraft}, R.~P., \& {Fulbright}, J. 1998, \aj,
  116, 1286

\bibitem[{{Hawkins} {et~al.}(2015){Hawkins}, {Jofr{\'e}}, {Masseron}, \&
  {Gilmore}}]{hawkins15}
{Hawkins}, K., {Jofr{\'e}}, P., {Masseron}, T., \& {Gilmore}, G. 2015, \mnras,
  453, 758

\bibitem[{{Hayes} {et~al.}(2018){Hayes}, {Majewski}, {Shetrone},
  {Fern{\'a}ndez-Alvar}, {Allende Prieto}, {Schuster}, {Carigi}, {Cunha},
  {Smith}, {Sobeck}, {Almeida}, {Beers}, {Carrera}, {Fern{\'a}ndez-Trincado},
  {Garc{\'\i}a-Hern{\'a}ndez}, {Geisler}, {Lane}, {Lucatello}, {Matthews},
  {Minniti}, {Nitschelm}, {Tang}, {Tissera}, \& {Zamora}}]{hayes18}
{Hayes}, C.~R., {Majewski}, S.~R., {Shetrone}, M., {et~al.} 2018, \apj, 852, 49

\bibitem[{{Haywood} {et~al.}(2018){Haywood}, {Di Matteo}, {Lehnert}, {Snaith},
  {Fragkoudi}, \& {Khoperskov}}]{haywood18}
{Haywood}, M., {Di Matteo}, P., {Lehnert}, M., {et~al.} 2018, \aap, 618, A78

\bibitem[{{Helmi}(2020)}]{helmi20}
{Helmi}, A. 2020, \araa, 58, 205

\bibitem[{{Helmi} {et~al.}(2018){Helmi}, {Babusiaux}, {Koppelman}, {Massari},
  {Veljanoski}, \& {Brown}}]{helmi18}
{Helmi}, A., {Babusiaux}, C., {Koppelman}, H.~H., {et~al.} 2018, \nat, 563, 85

\bibitem[{{Helmi} {et~al.}(1999){Helmi}, {White}, {de Zeeuw}, \&
  {Zhao}}]{helmi99}
{Helmi}, A., {White}, S. D.~M., {de Zeeuw}, P.~T., \& {Zhao}, H. 1999, \nat,
  402, 53

\bibitem[{{Honda} {et~al.}(2011){Honda}, {Aoki}, {Beers}, \&
  {Takada-Hidai}}]{honda11}
{Honda}, S., {Aoki}, W., {Beers}, T.~C., \& {Takada-Hidai}, M. 2011, \apj, 730,
  77

\bibitem[{{Ishigaki} {et~al.}(2010){Ishigaki}, {Chiba}, \& {Aoki}}]{ishigaki10}
{Ishigaki}, M., {Chiba}, M., \& {Aoki}, W. 2010, \pasj, 62, 143

\bibitem[{{Ishigaki} {et~al.}(2013){Ishigaki}, {Aoki}, \& {Chiba}}]{ishigaki13}
{Ishigaki}, M.~N., {Aoki}, W., \& {Chiba}, M. 2013, \apj, 771, 67

\bibitem[{{Ishigaki} {et~al.}(2012){Ishigaki}, {Chiba}, \& {Aoki}}]{ishigaki12}
{Ishigaki}, M.~N., {Chiba}, M., \& {Aoki}, W. 2012, \apj, 753, 64

\bibitem[{{Ivans} {et~al.}(2003){Ivans}, {Sneden}, {James}, {Preston},
  {Fulbright}, {H{\"o}flich}, {Carney}, \& {Wheeler}}]{ivans03}
{Ivans}, I.~I., {Sneden}, C., {James}, C.~R., {et~al.} 2003, \apj, 592, 906

\bibitem[{{Iwamoto} {et~al.}(1999){Iwamoto}, {Brachwitz}, {Nomoto},
  {Kishimoto}, {Umeda}, {Hix}, \& {Thielemann}}]{iwamoto99}
{Iwamoto}, K., {Brachwitz}, F., {Nomoto}, K., {et~al.} 1999, \apjs, 125, 439

\bibitem[{{Jean-Baptiste} {et~al.}(2017){Jean-Baptiste}, {Di Matteo},
  {Haywood}, {G{\'o}mez}, {Montuori}, {Combes}, \& {Semelin}}]{jean-baptiste17}
{Jean-Baptiste}, I., {Di Matteo}, P., {Haywood}, M., {et~al.} 2017, \aap, 604,
  A106

\bibitem[{{Jonsell} {et~al.}(2005){Jonsell}, {Edvardsson}, {Gustafsson},
  {Magain}, {Nissen}, \& {Asplund}}]{jonsell05}
{Jonsell}, K., {Edvardsson}, B., {Gustafsson}, B., {et~al.} 2005, \aap, 440,
  321

\bibitem[{{Kobayashi} {et~al.}(2006){Kobayashi}, {Umeda}, {Nomoto}, {Tominaga},
  \& {Ohkubo}}]{kobayashi06}
{Kobayashi}, C., {Umeda}, H., {Nomoto}, K., {Tominaga}, N., \& {Ohkubo}, T.
  2006, \apj, 653, 1145

\bibitem[{{Koppelman} {et~al.}(2020){Koppelman}, {Bos}, \&
  {Helmi}}]{koppelman20}
{Koppelman}, H.~H., {Bos}, R. O.~Y., \& {Helmi}, A. 2020, \aap, 642, L18

\bibitem[{{Koppelman} {et~al.}(2019{\natexlab{a}}){Koppelman}, {Helmi},
  {Massari}, {Price-Whelan}, \& {Starkenburg}}]{koppelman19a}
{Koppelman}, H.~H., {Helmi}, A., {Massari}, D., {Price-Whelan}, A.~M., \&
  {Starkenburg}, T.~K. 2019{\natexlab{a}}, \aap, 631, L9

\bibitem[{{Koppelman} {et~al.}(2019{\natexlab{b}}){Koppelman}, {Helmi},
  {Massari}, {Roelenga}, \& {Bastian}}]{koppelman19b}
{Koppelman}, H.~H., {Helmi}, A., {Massari}, D., {Roelenga}, S., \& {Bastian},
  U. 2019{\natexlab{b}}, \aap, 625, A5

\bibitem[{{Korotin} {et~al.}(2015){Korotin}, {Andrievsky}, {Hansen}, {Caffau},
  {Bonifacio}, {Spite}, {Spite}, \& {Fran{\c c}ois}}]{korotin15}
{Korotin}, S.~A., {Andrievsky}, S.~M., {Hansen}, C.~J., {et~al.} 2015, \aap,
  581, A70

\bibitem[{{Lach} {et~al.}(2020){Lach}, {R{\"o}pke}, {Seitenzahl}, {Cot{\'e}},
  {Gronow}, \& {Ruiter}}]{lach20}
{Lach}, F., {R{\"o}pke}, F.~K., {Seitenzahl}, I.~R., {et~al.} 2020, \aap, 644,
  A118

\bibitem[{{Laird} {et~al.}(1988){Laird}, {Carney}, \& {Latham}}]{laird88}
{Laird}, J.~B., {Carney}, B.~W., \& {Latham}, D.~W. 1988, \aj, 95, 1843

\bibitem[{{Lind} {et~al.}(2011){Lind}, {Asplund}, {Barklem}, \&
  {Belyaev}}]{lind11}
{Lind}, K., {Asplund}, M., {Barklem}, P.~S., \& {Belyaev}, A.~K. 2011, \aap,
  528, A103

\bibitem[{{Lindegren} {et~al.}(2020){Lindegren}, {Bastian}, {Biermann},
  {Bombrun}, {de Torres}, {Gerlach}, {Geyer}, {Hern{\'a}ndez}, {Hilger},
  {Hobbs}, {Klioner}, {Lammers}, {McMillan}, {Ramos-Lerate},
  {Steidelm{\"u}ller}, {Stephenson}, \& {van Leeuwen}}]{lindegren20}
{Lindegren}, L., {Bastian}, U., {Biermann}, M., {et~al.} 2020, arXiv e-prints,
  arXiv:2012.01742

\bibitem[{{Lodders}(2003)}]{lodders03}
{Lodders}, K. 2003, \apj, 591, 1220

\bibitem[{{Luyten}(1979)}]{luyten79}
{Luyten}, W.~J. 1979, {NLTT catalogue. Volume\_I. +90\_\_to\_+30\_.
  Volume.\_II. +30\_\_to\_0\_.} (Univ. Minnesota, Minneapolis)

\bibitem[{{Mackereth} {et~al.}(2019){Mackereth}, {Schiavon}, {Pfeffer},
  {Hayes}, {Bovy}, {Anguiano}, {Allende Prieto}, {Hasselquist}, {Holtzman},
  {Johnson}, {Majewski}, {O'Connell}, {Shetrone}, {Tissera}, \&
  {Fern{\'a}ndez-Trincado}}]{mackereth19}
{Mackereth}, J.~T., {Schiavon}, R.~P., {Pfeffer}, J., {et~al.} 2019, \mnras,
  482, 3426

\bibitem[{{Mashonkina} {et~al.}(2017){Mashonkina}, {Sitnova}, \&
  {Belyaev}}]{mashonkina17}
{Mashonkina}, L., {Sitnova}, T., \& {Belyaev}, A.~K. 2017, \aap, 605, A53

\bibitem[{{Matsuno} {et~al.}(2019){Matsuno}, {Aoki}, \& {Suda}}]{matsuno19}
{Matsuno}, T., {Aoki}, W., \& {Suda}, T. 2019, \apjl, 874, L35

\bibitem[{{McMillan}(2017)}]{mcmillan17}
{McMillan}, P.~J. 2017, \mnras, 465, 76

\bibitem[{{McWilliam}(1997)}]{mcwilliam97}
{McWilliam}, A. 1997, \araa, 35, 503

\bibitem[{{Mel{\'e}ndez} {et~al.}(2009){Mel{\'e}ndez}, {Asplund}, {Gustafsson},
  \& {Yong}}]{melendez09}
{Mel{\'e}ndez}, J., {Asplund}, M., {Gustafsson}, B., \& {Yong}, D. 2009, \apjl,
  704, L66

\bibitem[{{Monty} {et~al.}(2020){Monty}, {Venn}, {Lane}, {Lokhorst}, \&
  {Yong}}]{monty20}
{Monty}, S., {Venn}, K.~A., {Lane}, J. M.~M., {Lokhorst}, D., \& {Yong}, D.
  2020, \mnras, 497, 1236

\bibitem[{{Nagasawa} {et~al.}(2018){Nagasawa}, {Marshall}, {Li}, {Hansen},
  {Simon}, {Bernstein}, {Balbinot}, {Drlica-Wagner}, {Pace}, {Strigari},
  {Pellegrino}, {DePoy}, {Suntzeff}, {Bechtol}, {Walker}, {Abbott}, {Abdalla},
  {Allam}, {Annis}, {Benoit-L{\'e}vy}, {Bertin}, {Brooks}, {Carnero Rosell},
  {Carrasco Kind}, {Carretero}, {Cunha}, {D'Andrea}, {da Costa}, {Davis},
  {Desai}, {Doel}, {Eifler}, {Flaugher}, {Fosalba}, {Frieman},
  {Garc{\'\i}a-Bellido}, {Gaztanaga}, {Gerdes}, {Gruen}, {Gruendl}, {Gschwend},
  {Gutierrez}, {Hartley}, {Honscheid}, {James}, {Jeltema}, {Krause}, {Kuehn},
  {Kuhlmann}, {Kuropatkin}, {March}, {Miquel}, {Nord}, {Roodman}, {Sanchez},
  {Santiago}, {Scarpine}, {Schindler}, {Schubnell}, {Sevilla-Noarbe}, {Smith},
  {Smith}, {Soares-Santos}, {Sobreira}, {Suchyta}, {Tarle}, {Thomas}, {Tucker},
  {Wechsler}, {Wolf}, \& {Yanny}}]{nagasawa18}
{Nagasawa}, D.~Q., {Marshall}, J.~L., {Li}, T.~S., {et~al.} 2018, \apj, 852, 99

\bibitem[{{Nissen} {et~al.}(2007){Nissen}, {Akerman}, {Asplund}, {Fabbian},
  {Kerber}, {Kaufl}, \& {Pettini}}]{nissen07}
{Nissen}, P.~E., {Akerman}, C., {Asplund}, M., {et~al.} 2007, \aap, 469, 319

\bibitem[{{Nissen} {et~al.}(2014){Nissen}, {Chen}, {Carigi}, {Schuster}, \&
  {Zhao}}]{nissen14}
{Nissen}, P.~E., {Chen}, Y.~Q., {Carigi}, L., {Schuster}, W.~J., \& {Zhao}, G.
  2014, \aap, 568, A25

\bibitem[{{Nissen} \& {Schuster}(1997)}]{nissen97}
{Nissen}, P.~E. \& {Schuster}, W.~J. 1997, \aap, 326, 751

\bibitem[{{Nissen} \& {Schuster}(2010)}]{nissen10}
{Nissen}, P.~E. \& {Schuster}, W.~J. 2010, \aap, 511, L10

\bibitem[{{Nissen} \& {Schuster}(2011)}]{nissen11}
{Nissen}, P.~E. \& {Schuster}, W.~J. 2011, \aap, 530, A15

\bibitem[{{Nomoto} {et~al.}(2013){Nomoto}, {Kobayashi}, \&
  {Tominaga}}]{nomoto13}
{Nomoto}, K., {Kobayashi}, C., \& {Tominaga}, N. 2013, \araa, 51, 457

\bibitem[{{North} {et~al.}(2012){North}, {Cescutti}, {Jablonka}, {Hill},
  {Shetrone}, {Letarte}, {Lemasle}, {Venn}, {Battaglia}, {Tolstoy}, {Irwin},
  {Primas}, \& {Fran{\c{c}}ois}}]{north12}
{North}, P., {Cescutti}, G., {Jablonka}, P., {et~al.} 2012, \aap, 541, A45

\bibitem[{{Ram{\'\i}rez} {et~al.}(2019){Ram{\'\i}rez}, {Khanal}, {Lichon},
  {Chanam{\'e}}, {Endl}, {Mel{\'e}ndez}, \& {Lambert}}]{ramirez19}
{Ram{\'\i}rez}, I., {Khanal}, S., {Lichon}, S.~J., {et~al.} 2019, \mnras, 490,
  2448

\bibitem[{{Ram{\'\i}rez} {et~al.}(2012){Ram{\'\i}rez}, {Mel{\'e}ndez}, \&
  {Chanam{\'e}}}]{ramirez12}
{Ram{\'\i}rez}, I., {Mel{\'e}ndez}, J., \& {Chanam{\'e}}, J. 2012, \apj, 757,
  164

\bibitem[{{Reggiani} \& {Mel{\'e}ndez}(2018)}]{reggiani18}
{Reggiani}, H. \& {Mel{\'e}ndez}, J. 2018, \mnras, 475, 3502

\bibitem[{{Revaz} {et~al.}(2009){Revaz}, {Jablonka}, {Sawala}, {Hill},
  {Letarte}, {Irwin}, {Battaglia}, {Helmi}, {Shetrone}, {Tolstoy}, \&
  {Venn}}]{revaz09}
{Revaz}, Y., {Jablonka}, P., {Sawala}, T., {et~al.} 2009, \aap, 501, 189

\bibitem[{{Roederer} {et~al.}(2010){Roederer}, {Sneden}, {Thompson}, {Preston},
  \& {Shectman}}]{roederer10}
{Roederer}, I.~U., {Sneden}, C., {Thompson}, I.~B., {Preston}, G.~W., \&
  {Shectman}, S.~A. 2010, \apj, 711, 573

\bibitem[{{Ryan}(1992)}]{ryan92}
{Ryan}, S.~G. 1992, \aj, 104, 1144

\bibitem[{{Sch{\"o}nrich} {et~al.}(2010){Sch{\"o}nrich}, {Binney}, \&
  {Dehnen}}]{schoenrich10}
{Sch{\"o}nrich}, R., {Binney}, J., \& {Dehnen}, W. 2010, \mnras, 403, 1829

\bibitem[{{Schuster} {et~al.}(2012){Schuster}, {Moreno}, {Nissen}, \&
  {Pichardo}}]{schuster12}
{Schuster}, W.~J., {Moreno}, E., {Nissen}, P.~E., \& {Pichardo}, B. 2012, \aap,
  538, A21

\bibitem[{{Schuster} {et~al.}(1993){Schuster}, {Parrao}, \& {Contreras
  Martinez}}]{schuster93}
{Schuster}, W.~J., {Parrao}, L., \& {Contreras Martinez}, M.~E. 1993, \aaps,
  97, 951

\bibitem[{{Scott} {et~al.}(2015){Scott}, {Asplund}, {Grevesse}, {Bergemann}, \&
  {Sauval}}]{scott15}
{Scott}, P., {Asplund}, M., {Grevesse}, N., {Bergemann}, M., \& {Sauval}, A.~J.
  2015, \aap, 573, A26

\bibitem[{{Sk{\'u}lad{\'o}ttir} {et~al.}(2017){Sk{\'u}lad{\'o}ttir}, {Tolstoy},
  {Salvadori}, {Hill}, \& {Pettini}}]{skuladottir17}
{Sk{\'u}lad{\'o}ttir}, {\'A}., {Tolstoy}, E., {Salvadori}, S., {Hill}, V., \&
  {Pettini}, M. 2017, \aap, 606, A71

\bibitem[{{Stephens} \& {Boesgaard}(2002)}]{stephens02}
{Stephens}, A. \& {Boesgaard}, A.~M. 2002, \aj, 123, 1647

\bibitem[{{Takeda} {et~al.}(2005){Takeda}, {Hashimoto}, {Taguchi}, {Yoshioka},
  {Takada-Hidai}, {Saito}, \& {Honda}}]{takeda05b}
{Takeda}, Y., {Hashimoto}, O., {Taguchi}, H., {et~al.} 2005, \pasj, 57, 751

\bibitem[{{Tolstoy} {et~al.}(2009){Tolstoy}, {Hill}, \& {Tosi}}]{tolstoy09}
{Tolstoy}, E., {Hill}, V., \& {Tosi}, M. 2009, \araa, 47, 371

\bibitem[{{Umeda} \& {Nomoto}(2002)}]{umeda02}
{Umeda}, H. \& {Nomoto}, K. 2002, \apj, 565, 385

\bibitem[{{Venn} {et~al.}(2012){Venn}, {Shetrone}, {Irwin}, {Hill}, {Jablonka},
  {Tolstoy}, {Lemasle}, {Divell}, {Starkenburg}, {Letarte}, {Baldner},
  {Battaglia}, {Helmi}, {Kaufer}, \& {Primas}}]{venn12}
{Venn}, K.~A., {Shetrone}, M.~D., {Irwin}, M.~J., {et~al.} 2012, \apj, 751, 102

\bibitem[{{Wilhelm} {et~al.}(1999){Wilhelm}, {Beers}, {Sommer-Larsen}, {Pier},
  {Layden}, {Flynn}, {Rossi}, \& {Christensen}}]{wilhelm99}
{Wilhelm}, R., {Beers}, T.~C., {Sommer-Larsen}, J., {et~al.} 1999, \aj, 117,
  2329

\bibitem[{{Yan} {et~al.}(2016){Yan}, {Shi}, {Nissen}, \& {Zhao}}]{yan16}
{Yan}, H.~L., {Shi}, J.~R., {Nissen}, P.~E., \& {Zhao}, G. 2016, \aap, 585,
  A102

\end{thebibliography}

\end{document}